\documentstyle[]{article}

\begin{document}

\title{Geometric Formulation of Nonlinear Quantum Mechanics for Density
        Matrices\thanks{The talk presented at the conference: ``New
Insights in Quantum Mechanics'', Goslar, Aug. 31st -- Sept. 4th, 1998.}}

\author{Pavel B\'ona\\
Department of Theoretical Physics, Faculty of Mathematics and
Physics\\ Comenius University, SK-842 15 Bratislava, Slovakia
\\E-mail: bona@sophia.dtp.fmph.uniba.sk}
%%%%%%%%%%%%%%%%%%%%%%%%%%%%%%%%%%%%%%%%%%%%%%%%%%%%%%%%%%%%%%%%%%%%%%%%%%
%%%\input{98-mac}%%%%%%%%%%%%%%%%%%%%

\hyphenation{co-ad-joint re-pre-sen-ta-ti-on Bra-ti-sla-va}

\def\nl{\hfil\break}
\def\acc{$\!\!$\'{}$\!$}
\def\dti{\!\cdot\!}
\def\rref#1~{~(\ref{eq;#1})}

\def\rhb{$\{\mrh\}$}
\def\mrhb{\{\mrh\}}
\def\mSs{{\cal S}_*}
\def\Ss{${\cal S}_*$}
\def\CTs{$C^\infty(\mcl T_s(\mH),\mbR)$}
\def\mCTs{C^\infty(\mcl T_s(\mH),\mbR)}

\def\refer#1#2#3#4#5{#1:\ {\sl #2}\ {\bf #3}\ {(#4)}\ #5;\ }

\def\bC{$\Bbb C$}
\def\mbC{{\Bbb C}}
\def\bR{$\Bbb R$}
\def\mbR{{\Bbb R}}
\def\bN{$\Bbb N$}
\def\mbN{{\Bbb N}}
\def\iu{{\it u}}

\def\rf{{\rm f}}
\def\rx{{\rm x}}
\def\ry{{\rm y}}
\def\rz{{\rm z}}
\def\ra{{\rm a}}
\def\rrb#1{{\rm b}_{#1}}
\def\rc{{\rm c}}
\def\rd{{\rm d}}
\def\rQ{{\rm Q}}
\def\ru{{\rm u}}
\def\rS{{\rm S}}

\def\Lqn{$L^2(\mbR^n)$}
\def\mLqn{L^2(\mbR^n)}
\def\LHs{${\cal L(H)}_s$}
\def\mLHs{{\cal L(H)}_s}
\def\mH{{\cal H}}
\def\mK{{\cal K}}
\def\mLH{{\cal L(H)}}
\def\H{${\cal H}$}
\def\K{${\cal K}$}
\def\PH{$P({\cal H})$}
\def\mPH{P({\cal H})}
\def\LH{${\cal L(H)}$}
\def\A{${\cal A}$}
\def\mA{{\cal A}}
\def\Z{${\cal Z}$}
\def\mZ{{\cal Z}}
\def\C{${\cal C}$}
\def\Cc{${\cal C}_{cl}$}
\def\mCc{{\cal C}_{cl}}
\def\mC{{\cal C}}
\def\B{${\cal B}$}
\def\mB{{\cal B}}
\def\F{${\cal F}$}
\def\mF{{\cal F}}
\def\Fr{$\mF_{\mrh}$}
\def\mFr{\mF_{\mrh}}

\def\cl#1{${\cal #1}$}
\def\mcl#1{{\cal #1}}

\def\P#1{$P_{#1}$}
\def\mP#1{P_{#1}}

\def\D#1,#2~{$D_{#2}{#1}$}
\def\mD#1,#2~{D_{#2}{#1}}
\def\Dfr{\D f,\mrh~}
\def\dfr{\d f,\mrh~}
\def\mDfr{\mD f,\mrh~}
\def\mdfr{\md f,\mrh~}
\def\Dfn{\D f,\nu~}
\def\dfn{\d f,\nu~}
\def\mDfn{\mD f,\nu~}
\def\mdfn{\md f,\nu~}
\def\Dhr{\D h,\mrh~}
\def\dhr{\d h,\mrh~}
\def\mDhr{\mD h,\mrh~}
\def\mdhr{\md h,\mrh~}
\def\mDhn{\mD h,\nu~}
\def\mdhn{\md h,\nu~}
\def\Dhn{\D h,\nu~}
\def\dhn{\d h,\nu~}

\def\veps{$\varepsilon$\ }
\def\eps{$\epsilon$\ }
\def\mveps{\varepsilon}
\def\meps{\epsilon}
\def\alp{$\alpha$}
\def\malp{\alpha}
\def\mphi{{\varphi}}
\def\rh{$\varrho$}
\def\mrh{\varrho}
\def\mlam{\lambda}
\def\lam{$\lambda$}

\def\mh#1{{\rm h}_{#1}}
\def\h#1{${\rm h}_{#1}$}

\def\ph#1,#2~{$\varphi_{#1}^{#2}$}
\def\mph#1,#2~{\varphi_{#1}^{#2}}
\def\cmrh#1~{{\varrho_{#1}}}
\def\crh#1~{$\varrho_{#1}$\ }

\def\pph#1,#2~{$\tilde\mph#1,#2~$}
\def\mpph#1,#2~{\tilde\mph#1,#2~}
\def\un#1,#2,#3~{${\rm u}_#1(#2,#3)$}
\def\mun#1,#2,#3~{{\rm u}_#1(#2,#3)}
\def\gQ#1,#2~{$g_\rQ(#1,#2)$}
\def\mgQ#1,#2~{g_\rQ(#1,#2)}
\def\taQ{$\tau^\rQ$}
\def\mtaQ{\tau^\rQ}
\def\mtQ#1,#2~{\tau^\rQ_{#1}#2}
\def\tQ#1,#2~{$\tau^\rQ_{#1}#2$}

\def\Lq{$L^2({\Bbb R},dq)$}
\def\mLq{L^2({\Bbb R},dq)}

\def\Ca{$C^*$-algebra}
\def\Csa{$C^*$-subalgebra}
\def\autm{${}^*$-automorphism}
\def\aut#1{${}^*$-Aut\ #1}
\def\maut#1{{}^*$-Aut$\ #1}

\def\lb{\langle}
\def\rb{\rangle}

\def\eequiv{\Leftrightarrow}
\def\imply{\Rightarrow}

\def\wrt{with respect to\ }

\def\prob{{\rm prob}}
\def\nbhd{neighbourhood\ }

\def\noidt{\noindent}

%% Some other macros used in the sample text
\def\st{\scriptstyle}
\def\sst{\scriptscriptstyle}
\def\mco{\multicolumn}
\def\epp{\epsilon^{\prime}}
\def\vep{\varepsilon}
\def\ppg{\pi^+\pi^-\gamma}
\def\vp{{\bf p}}
\def\ko{K^0}
\def\kb{\bar{K^0}}
\def\al{\alpha}
\def\ab{\bar{\alpha}}
\def\be{\begin{equation}}
\def\ee{\end{equation}}
\def\bea{\begin{eqnarray}}
\def\eea{\end{eqnarray}}
\def\CPbar{\hbox{{\rm CP}\hskip-1.80em{/}}}%temp replacement due to no font

\newcommand{\bequ}{\begin{equation}}
\newcommand{\enqu}{\end{equation}}
\newcommand{\barr}{{\begin{eqnarray}}}
\newcommand{\earr}{{\end{eqnarray}}}

\newcommand{\rarw}{\rightarrow}
\newcommand{\Bbb}{\bf}

\newtheorem{thm}{Theorem}
%[section]
\newtheorem{pt}[thm]{}
\newtheorem{rem}{{Remark}}
%[thm]
\newtheorem{exmp}{\rm Examples}
\newtheorem{exm}{Example}
\newtheorem{note}[thm]{Notes}

%%%%%%%%%%%%%%%%%%%%%%%%%%%%%%%%%%%%%%%%%%%%%%%%%%%%%%%%%%%%%%%%%%

\maketitle

\begin{abstract}{
Proposals for nonlinear extenstions
of quantum mechanics are discussed. Two different
concepts of ``mixed state'' for any nonlinear
version of quantum theory are introduced: (i) {\em genuine mixture}
corresponds to operational ``mixing'' of different ensembles,
 and (ii) a mixture described by single density matrix without
having a canonical operational possibility to pick out its specific convex
decomposition is called here an {\em elementary
mixture}. Time evolution of a class of nonlinear extensions of quantum
mechanics is introduced. Evolution of an elementary mixture cannot
be generally given by evolutions of components of its arbitrary convex
decompositions.
The theory is formulated in a ``geometric form'': It can be considered as
a version of Hamiltonian mechanics on infinite dimensional space of density
matrices. A quantum interpretation of the theory is sketched.}
\end{abstract}

\section{Introduction}
In popular and well written book~\cite{peres} on conceptual foundations of
quantum mechanics (QM) there is a subsection [Chap. 9-4, p.278] on nonlinear
Schr\"odinger equation
inserted into the section entitled {\em Some impossible processes},
and containing a proof of inconsistency of any nonlinear evolution in QM ``if
we retain the {\em other postulates} of quantum theory without any
change''. These ``other postulates'' include statistical interpretation and
time evolution
of density matrices corresponding to some of our {\em genuine mixtures},
as well as the definition of entropy by von Neumann
which is appropriate just to our
{\em elementary mixtures} (cf. Subsection~\ref{subsec;mix}).
 The nonlinear extensions of QM appearing usually in
literature~\cite{bial&myc,weinb,doeb&gold} are formulated for time
evolution of vector states (i.e. wave functions); trials to extend such
dynamics to evolutions of density matrices led to inconsistencies, cf. e.g.
criticism~\cite{gisin} of~\cite{weinb}.

We argue that (at least some of) the
inconsistencies mentioned above are consequences of improper
interpretation of ``mixtures'' and of definition of their time--evolution.
We shall introduce two concepts of ``mixed states'' in this contribution,
as well as a class of nonlinear extensions of dynamics avoiding the mentioned
inconsistencies. Our formulation of nonlinear quantum theory~\cite{bona} was
inspired by symplectic reformulation of QM~\cite{bona1}, and was encouraged by
publication of the Weinberg's proposal~\cite{weinb}.

\section{Two concepts of ``mixture'' and nonlinearity in QM}\label{sec;mix}
In standard elementary formulation of QM (linear, without superselection rules,
describing systems with ``finite number of degrees of freedom'') the set of
bounded ``observables'' is described by the set \LHs\ of all bounded
selfadjoint operators in a complex Hilbert space \H, and the set of
``states'' is described by the set of density matrices $\mSs:=\mcl T(\mH)_{+1}$\
consisting of positive trace--class operators of unit trace.
Any density matrix $\mrh\in\mSs$\ is either of the form
\bequ\label{eq;1}
\mrh=P_\mphi=P_\mphi^2=P_\mphi^*\in\mLHs,\quad\mphi\in\mH,
\enqu
corresponding to a ``pure state'', or it is a nontrivial convex combination
of one dimensional projections $P_\mphi$.
 Each density matrix corresponding to
a nontrivial mixture, i.e. $\mrh^2\neq\mrh$, can be written in infinitely
many mutually different ways in the form of its convex decompositions:
\bequ\label{eq;2}
\mrh=\sum_j\mlam_j\mrh_j \left(=\sum_k\kappa_k\nu_k\right),\ \mrh_j,\nu_k\in\mcl
T(\mH)_{+1},\ \mlam_j,\kappa_k\in[0,1)\subset\mbR_+.
\enqu
The density matrices of the
form~(\ref{eq;2}) represent~\cite{peres} ``mixed states'', resp. ``mixtures''.

 Each such a state is uniquelly
determined by determination of {\em expectations of ``sufficiently many''
observables} $A\in\mLHs$, i.e. by numbers
\bequ\label{eq;3}
\lb A\rb_\mrh:=Tr(\mrh\dti A),\quad\forall A\in\mLHs.
\enqu
Such a set of state--determining expectations corresponding to a
nontrivial mixture~(\ref{eq;2}) can arise in QM {\bf in two different
experimental situations} corresponding to {\em two possible
interpretations} of density matrices, cf. Subsection~\ref{subsec;mix}.
These two possibilities are not empirically distinguishable in the
standard framework of
QM, but their distinction becomes crucial, as we shall show in this paper,
in any nonlinear generalization of the theory.

\subsection{Two concepts of mixed states}\label{subsec;mix}
Let us distinguish here two kinds of preparation procedures for states
described in QM by a density matrix \rh, and let us also introduce the
corresponding two concepts of ``mixed states''.\nl
\noidt\ (i).\ \ A state of a given system in QM described by a
density matrix $\mrh\neq\mrh^2$\ is often interpreted as representation
of the statistical ensemble \rhb\ of copies of the considered physical system
consisting of subensembles numbered by an index $j\in J(\equiv$\ {\sl an
index set}), each of which is prepared by its own macroscopicaly
distinguishable preparation procedure preparing the system in a state
$\mrh_j\in\mSs$, and occuring in \rhb\ with probability $\mlam_j$,
$\mrh=\sum_j\mlam_j\mrh_j$. Hence the term ``mixture'' corresponds here to
intuitive notion of ``mixing'' of a set of copies of the system occuring in
states $\mrh_j$\ with relative frequencies $\mlam_j$. There are
infinitely many of such mutually different (possibly quantummechanicaly mutually
incompatible) preparation procedures leading to the same density matrix \rh\
which are in the framework of QM indistinguishable~\cite{peres} by
measurements of the quantities \rref 3~. Each of the
``components'' $\mrh_j$\ such that $\mrh_j\neq\mrh^2_j$\ could be decomposed
further into ``less mixed'' components, $\mrh_j=\sum_k\mlam_{jk}\mrh_{jk}$,
etc. Let the components $\mrh_j$\ of the decomposition corresponding to the
described preparation procedure of the ensemble \rhb\ be {\em empirically
indecomposable}, i.e. they are {\em elementary mixtures} in the sense of the
next point (ii). Then the state corresponding to the ensemble \rhb\ will be
called a {\bf genuine mixture}. It is characterized by a specific convex
combination of elementary mixtures:
$\mrhb\equiv\mrhb_\mlam:=\{(\mlam_j;\mrh_j):j\in J\}$. Other such
decompositions of the same density matrix will correspond to different genuine
mixtures (empirically distinguishable in some generalizations of QM, cf.
later in this paper).

\noidt(ii).\ \ Let a density matrix $\mrh\in\mSs$\  associated with the
considered system \rS\ be given. It is always possible (at least
mathematically) to find another QM--system $\rS'$\ and a vector state
$\Phi\in\mH\otimes\mH'$\ of the combined system $\rS+\rS'$\ such that
its restriction to the subsystem \rS\ is \rh:
\bequ\label{eq;restric}
Tr(P_\Phi\dti A\otimes I_{\mH'})\equiv Tr(\mrh\dti A),\quad\forall A\in\mLHs,
\enqu
where $I_{\mH'}\in\mcl L(\mH')$\ is the unit observable of the attached system
$\rS'$ (called sometimes an ``ancilla'').

Let us assume that the state \rh\ of \rS\ is obtained by such a restriction
from an empirically prepared state $P_\Phi$\ of a system $\rS+\rS'$, the
observables of which contain all elements of $\mcl{L(H\otimes H')}_s$\
(hence, no superselection rules). The systems \rS\
and $\rS'$\ are {\em dynamically
independent}. The state \rh\ of \rS\ (and the corresponding
statistical ensemble of copies of \rS) prepared in this way will be called
an {\bf elementary mixture}. These states are a special case of genuine
mixtures corresponding to the trivial decomposition of \rh\ into a single
element.

This second method of preparation of a state \rh\ do not specify any
decomposition of \rh\ into ``simpler'', or ``purer'' states.
We need not specify in the following a way of preparation of an elementary
mixture; in the forthcoming considerations, {\em any elementary mixture
\rh\ will be a ``full fledged'' description of a quantum state}, in an equal
degree for nontrivial mixtures $\mrh^2\neq\mrh$, as well as for pure
states. From the point of view of physical intuition, the elementary
mixtures are distinguished by impossibility to specify their decomposition
to subensembles {\em determined by their physical preparation}.
\begin{rem}
We shall see how nonlinear quantum dynamics can distinguish
different genuine mixtures with the same ``barycentre'' \rh\ by measuring
the quantities in \rref 3~ only.
Detailed specification of conditions under which
we are dealing either with ``genuine'', or with ``elementary'' mixture will
be actual only after observing some nonlinear quantum evolution: It is
also a question for experimentalists. Here we stress just theoretical
possibilities and their connections.
\end{rem}
These different possible interpretations of density matrices are usually
ignored because of their {\em experimental indistinguishability} in
(linear) QM.\footnote{The only situations where
the present author noticed a discussion of these different interpretations
 were works on {\em quantum
measurement problem}~\cite{B&L&M}. The genuine mixtures are called there
also {\em Gemenge}, and the interpretation: {\em ignorance
interpretation}.}
 They could
be distinguished by considering also states of preparation apparatuses, i.e.
by considering correlations of states of \rS\ with some ``macroscopic
parameters'' (we shall not go into details of this point here, cf., however,
Remark~\ref{rem;obs}).
\begin{rem}\label{rem;obs}
Te above physically intuitive determination of difference between the two
concepts of ``mixed states'' can be made mathematically clear after
accepting an extended set of ``observables'' for the considered system \rS.
Let the \Ca\ of observables be the set $C_s(\overline\mSs,\mLH)$\ of continuous
(in some conveniently specified topologies) functions $\hat \ra:\nu\mapsto
\ra(\nu)\in\mLH$\ on ``conveniently'' compactified \Ss,
$\overline\mSs\ni\nu$, with bounded operator values. The centre of this
\Ca\ contains {\bf classical quantities} consisting of the commutative \Ca\
of scalar--valued continuous functions $a\in C(\overline\mSs)$. The
elementary mixtures are then just the pure states of this commutative subalgebra
.
Intuitively, this extension of the set of observables corresponds to the
extension of the physical system \rS\ by macroscopic
parameters of (preparation procedures of) its (microscopic) states.
\end{rem}

\subsection{Nonlinear transformations distinguish the two kinds of mixed
states}\label{subsec;nlin}
The above mentioned indistinguishability of different
decompositions~(\ref{eq;2}) of any
given mixture described by a density matrix \rh\ is conserved by linear
time evolutions (or any unitary transformations)
\bequ\label{eq;lin}
 \mphi_t(\mrh):=\ru_t\mrh \ru_t^*,\quad \ru_t\equiv\exp(-itH).
 \enqu
We have then for all $t\in\mbR$:
\bequ\label{eq;5}
\mphi_t(\mrh)=\sum_j\mlam_j\mphi_t(\mrh_j)=\sum_k\kappa_k\mphi_t(\nu_k),\
\mphi_{t=0}(\mrh)\equiv\mrh.
\enqu
Conversely, if a continuous group of invertible transformations $\mphi_t$\
of \Ss\ satisfies~(\ref{eq;5}) for all decompositions~(\ref{eq;2}) of all
density matrices \rh, then it is described by some unitary operators
$\ru_t\equiv\exp(-itH)$,\cite{bra&rob} [Theorem 3.2.8, and Example
3.2.14]. In these cases, determination of time
evolution for pure states is sufficient to determine evolution
of all states uniquely.

Let us assume that $\mphi_t$\ is a nonlinear family of transformations of
\Ss, i.e.~(\ref{eq;5}) is no more identically valid.
 Hence there is \rh\ and its decompositions~(\ref{eq;2})
such that for a ``nonlinear $t$'':
\bequ\label{eq;6}
\left(\mphi_t(\mrh)\neq\right)\ \sum_j\mlam_j\mphi_t(\mrh_j)\not\equiv
\sum_k\kappa_k\mphi_t(\nu_\kappa).
\enqu
In other words, the mappings $\mphi_t:\mSs\mapsto\mSs$\ are not all affine now,
and the evolution of components $\mrh_j$\ of a decomposition~(\ref{eq;2})
of a density matrix \rh\  need not
determine a ``corresponding'' evolution of \rh. Evolutions of elementary
mixtures have to be specified {\em on the whole space of quantum states}
\Ss. Moreover, eq.\rref 6~
shows that nonlinear evolution can distinguish between
different genuine mixtures {\em corresponding to the same initial density
matrix}.

\section{Nonlinear Extensions of Quantum Dynamics}\label{sec;dyn}
Let us specify here a class of evolutions generalizing the linear
ones~(\ref{eq;lin}). We shall ignore here problems with unboundedness of
generators. After defining an {\em evolution of all elementary
mixtures},\footnote{Similar dynamics for density matrices
was formulated by Czachor et al.~\cite{czachor} in terms of Lie--Nambu brackets.
}
we shall stress its distinction from the corresponding {\em evolutions for
genuine mixtures}. We shall mention briefly the statistical interpretation
of the generalized quantum theory. More details on the theory can be found
in~\cite{bona}.
\subsection{Dynamics on the quantum phase space}
Let us consider the space \Ss\ of all density matrices as the {\bf quantum
phase space}.
The dynamics on \Ss\ is just Hamiltonian (better: Poisson) classical
dynamics on the infinite--dimensional submanifold (with boundary)
\Ss\ of the linear space $\mcl T_s(\mH)\ (\equiv$ symmetric trace class)
endowed with the trace norm and with the Poisson brackets
\bequ\label{eq;Poiss}
\{f,h\}(\mrh)\equiv i\,Tr\left(\mrh[D_\mrh f,D_\mrh h]\right),\quad f,h\in
C^\infty(\mcl T_s(\mH),\mbR).
\enqu
Here $D_\mrh f$\ is the differential of $f\in C^\infty(\mcl T_s(\mH),\mbR)$\
considered (in a canonical way~\cite{bona}, cf. Remark~\ref{rem;oper})
as a bounded linear operator on
\H, and $[\cdot,\cdot]$\ is the commutator in \LH.
\begin{rem}\label{rem;oper}
The correspondence between \D f,\mrh~\ and operators in \LHs\ is given by the
duality between the space of selfadjoint trace class operators
$\mcl T_s(\mH)\ (\ni\mrh,\nu)$, and its dual \LHs\ ($\ni A,B$)\
representing (real--)linear functionals $\nu\mapsto\lb B;\nu\rb\in\mbR$ on
$\mcl T_s(\mH)$ so that $\lb B;\nu\rb:= Tr(B\dti\nu)$. Hence the differential
\D f,\mrh~\ calculated on the vector $\nu$\ according to its definition as
a linear functional on $\mcl T_s(\H)$, can be represented by the
operator $\tilde\mD f,\mrh~\in\mLHs$\ determined by the relation
\bequ\label{eq;oper}
\mD f,\mrh~(\nu)\equiv\lb\mD f,\mrh~;\nu\rb:= \left.\frac{d}{dt}
\right|_{t=0}f(\mrh+t\nu)\equiv Tr(\nu\dti\tilde\mD f,\mrh~).
\enqu
Being this valid for all $\nu\in\mcl T_s(\mH)$, the operator
$\tilde\mD f,\mrh~$\ is determined unambiguously by\rref oper~.
We shall use $\tilde D\equiv D$\ in the following.
\end{rem}
Let $Q\in\mCTs$ be a Hamiltonian for our dynamics (for ``more realistic''
cases this function $Q$\ is not everywhere defined and it is
unbounded)~\cite{bona}. Then the Hamiltonian flow
\ph t,Q~\ can be described~\cite{bona} by {\bf unitary cocycle} $\ru_Q$\
consisting of the set of unitary
operators $\{\mun Q,t,\mrh~:t\in\mbR,\mrh\in\mSs\}$ on \H\
satisfying a version of the {\bf nonlinear Schr\"odinger equation}:
\bequ\label{eq;nl-Sch}
i\,\frac{d}{dt}\mun Q,t,\mrh~=D_{\mrh(t)}Q\dti\mun Q,t,\mrh~,\quad
\mun Q,0,\mrh~\equiv0;
\enqu
  and fulfilling the {\em cocycle identity}:
\bequ\label{eq;cocycle}
\mun Q,t+s,\mrh~=\mun Q,s,{\mph t,Q~(\mrh)}~\mun
Q,t,\mrh~,\quad\forall s,t,\in\mbR.
\enqu
Here enters the nonlinearity via
\bequ\label{eq;evol}
 \mrh(t)\equiv\mph t,Q~(\mrh):=\mun Q,t,\mrh~\mrh\mun
 Q,t,\mrh~^*,\quad\mrh(0):=\mrh.
\enqu
The equation\rref nl-Sch~ leads to the evolution equation written directly
for $\mrh(t)$:
\bequ\label{eq;rho-t}
i\,\frac{d}{dt}\mrh(t)=[\mD Q,\mrh(t)~,\mrh(t)].
\enqu

Let us rewrite eq.\rref nl-Sch~\ into a common form of nonlinear
Schr\"odinger equation for wave
functions $\psi(t):=\mun Q,t,P_\psi~\psi\in\mH$\ (we set $D_\psi\equiv
D_{P_\psi}$):
\bequ
i\,\frac{d}{dt}\psi(t)=D_{\psi(t)}Q\dti\psi(t).
\enqu
It is seen from\rref evol~, and from unitarity of \un
Q,t,\mrh~\ that the evolution $\mphi^Q$\ leaves spectral characteristics of
all density matrices invariant. Hence it conserves also purity of states.
If the evolution were known for vector states only, evolution for density
matrices would remain undefined. In specific
cases, it is possible~\cite{czachor} to find a ``natural extension'' of the
Hamiltonian $Q$ from the subset $\mPH\subset\mSs$\ of vector states
to other parts of \Ss; mathematical guides for such an extension might
be, e.g. continuity, symmetry, or some aesthetical, resp. formal
considerations.

\begin{exm}\label{exm;nlin-weq}
Let us take, e.g. \H=\Lqn\ with
$\lb\psi|\mphi\rb:=\int\overline\psi(x)\mphi(x)\,d^nx$. Let us write
density matrices \rh\ ``in the $x$--representation'' with a help of their
operator kernels $\mrh(x,y)$:
\bequ\label{eq;kern}
 [\mrh\psi](x)\equiv\int\mrh(x,y)\psi(y)\,d^ny,\quad\psi\in\mH.
 \enqu
 Projection operators $P_\psi$ have the kernels
 $P_\psi(x,y)\equiv\|\psi\|^{-2}\psi(x)\overline\psi(y)$.
Let the Hamiltonian function $Q$\ will be taken as the (unbounded) functional
 \bequ\label{eq;QP}
 Q(P_\psi):= Tr(P_\psi\dti
H_0)+\frac{\mveps}{\malp+1}\int P_\psi(x,x)^{\malp+1}d^nx,
\enqu
with $H_0$\ some selfadjoint (linear) operator on \Lqn, and
$\malp>0$.\footnote{We shall proceed, in the presented ``Examples'', in a
heuristic way, by ``plausible'' formal manipulations; the necessary
mathematical comments are omitted here.}
Let $t\mapsto P_{\psi(t)}, \psi(0):=\psi$\ be any differentable curve
through $P_\psi\in\mPH$, and let $\dot{P}_\psi\in T_{P_\psi}\mPH$\ be its
tangent vector expressed by an operator according to Remark~\ref{rem;oper}.
Then the (unbounded, nonlinear) Hamiltonian \D Q,\psi~ can be expressed by:
\bequ
Tr(\mD
Q,\psi~\dti\dot{P}_\psi):=\left.\frac{d}{dt}\right|_{t=0}Q(P_{\psi(t)}),
\enqu
what leads to the corresponding form of
``nonlinear Schr\"odinger wave--equation'' for $\psi_t:=\psi(t)$:
\bequ\label{eq;nlin-weq}
i\,\left[\frac{d}{dt}\psi_t\right](x)=[H_0\psi_t](x)+\mveps
|\psi_t(x)|^{2\malp}\psi_t(x),\quad\|\psi_t\|\equiv 1.
\enqu
One possible extension of this nonlinear dynamics to the whole space \Ss\ is
obtained by ``the substitution $\mrh\mapsto P_\psi$'', i.e. by the choice of the
Hamiltonian
\bequ\label{eq;exten}
Q(\mrh):= Tr(\mrh\dti H_0)+\frac{\mveps}{\malp+1}\int\mrh(x,x)^{\malp+1}d^nx,
\enqu
and the corresponding dynamics is then described by\rref rho-t~ with
\bequ
\mD Q,\mrh~(\nu)\equiv Tr(\nu\dti H_0)+\mveps\int\mrh(x,x)^\malp\nu(x,x)\,d^nx.
\enqu
\end{exm}
We shall compare in the next subsection the evolutions of the mixed states
described by the same initial density
matrix\rref2~ for the two distinguished interpretations.

\subsection{Genuine mixtures, their dynamics and
interpretation}\label{subsec;mix-dyn}
 Since genuine mixtures $\mrhb:=\{\mrh\}_\mlam:=\{(\mlam_j;\mrh_j):\
 j\in J\}$\ are considered as
probability distributions over the space of ``elementary events'' \Ss\
(in the sense of the classical Kolmogorov probability theory)
with probabilities $\mlam_j$\ concentrated in the points
$\mrh_j\in\mSs$, adequate description of genuine mixtures are
probability measures $\mu_{\{\mrh\}}$\ on \Ss.
Let us denote $\delta_\mrh$\ the Dirac probability measures:
$\delta_\mrh(\Lambda)=0\eequiv\mrh\not\in\Lambda\subset\mSs$.
The measure $\mu_{\{\mrh\}}$\
describing the state--decomposition~(\ref{eq;2}) describes the integral of
functions $f\in L^1(\mSs,\mu_{\mrhb})$:
\bequ
\mu_{\{\mrh\}}(f)=\sum_j\mlam_j\delta_{\mrh_j}(f)=\sum_j\mlam_j f(\mrh_j).
\enqu
In this language, the
elementary mixtures might be considered as the special case of genuine
ones, in which the decomposition $\{\mrh\}$\ of \rh\ according to \rref 2~
is trivial:
$\mlam_1:=1,\ \mrh_1:=\mrh,\ \mlam_j\equiv0\ (\forall j\neq1)$.

If there is given a Hamiltonian time evolution \ph t,Q~\ on \Ss, the
evolution of the state described by $\mu_{\{\mrh\}}$\ can be described by the
corresponding evolution of the distinquished decomposition:
\bequ
\{\mrh\}:=\{(\mlam_j;\mrh_j):j\in J\}\mapsto\{\mrh\}_t:= \{(\mlam_j;\mph
t,Q~(\mrh_j)):j\in J\}.
\enqu
This evolution is described equivalently by the evolution of measures
$\mu_t$\ on \Ss:
\bequ
   (t;\mu)\mapsto \mu_t\equiv\mu\circ\mph -t,Q~.
\enqu
Let us illustrate, by explicit calculation, the difference between time
evolutions of the same initial
density matrix considered in its two different interpretations.
\begin{exm}
Let us take the system with its ``extended'' dynamics from
Example~\ref{exm;nlin-weq}, and let us fix a nontrivial
mixture \rh\ of several vector states $P_{\psi_j}$:
$\mrh=\sum_j\mlam_jP_{\psi_j}$. Let us calculate the difference between the
derivatives \wrt the time in
$t=0$\ of the two evolutions: (i) of the {\em genuine mixture evolution}
$\sum_j\mlam_j\mph t,Q~(P_{\psi_j})$, and (ii) of the {\em elementary mixture
evolution} $\mph t,Q~(\mrh)$.  We shall calculate the right side of\rref
rho-t~ for the two cases and take their difference. Let us write the kernel ``in
 $x$--representation'' of \rh\ as the convex
combination of the vector--state kernels:
\bequ
\mrh(x,y)\equiv\sum_j\mlam_j\|\psi_j\|^{-2}\psi_j(x)\overline\psi_j(y).
\enqu
The (symbolic) ``kernel'' of the Hamiltonian \D Q,\mrh~\ can be written:
\[ \mD Q,\mrh~(x,y)=H_0(x,y)+\mveps\delta(x-y)\mrh(x,x)^\malp. \]
Here, $\delta(\cdot)$\ is the Dirac distribution on $\mbR^n$. We have to
express the difference $\Delta^{\{\mrh\}}_t(x,y)$\
between the kernels (in x--representation) of the operators
\[ \sum_j\mlam_j[\mD Q,\psi_j(t)~,P_{\psi_j}],\ {\rm and}\ [\mD
Q,\mrh(t)~,\mrh(t)], \]
what expresses the difference between time derivatives of ``the
same density matrix'' $\mrh=\sum\mlam_jP_{\psi_j}$ in the two
interpretations. The linear operator $H_0$\ does not contribute into this
difference. The kernels of commutators entering into the calculation are
(for all $\nu\in\mSs$) of the form
\[ [\mD Q,\nu~,\nu](x,y)=[H_0,\nu](x,y)+\mveps\nu(x,y)(\nu(x,x)^\malp
-\nu(y,y)^\malp).\]
We can (and we shall) take all $\|\psi_j\|\equiv1$.
Let us denote
\[\chi_j^{\{\mrh\}}(x):=|\psi_j(x)|^{2\malp}-\left(\sum_k\mlam_k
|\psi_k(x)|^2\right)^\malp.\]
Then the wanted difference at $t=0$\ is
\bequ
\Delta_{\{\mrh\}}(x,y):=\Delta^{\{\mrh\}}_0(x,y)=\mveps\sum_j\mlam_j
\psi_j(x)\overline\psi_j(y)(\chi^{\{\mrh\}}_j(x)-\chi^{\{\mrh\}}_j(y)).
\enqu
By proving that the operator $\Delta_{\mrhb}$\ is not identical zero for all
$\{\mrh\}$,
we can prove
 nontrivial difference of the two time evolutions explicitly. This can be
easily proved for $\mlam_1:=1-\mlam_2$, and $\psi_1,\psi_2$\ specific
two--valued functions concentrated on disjoint subsets of $\mbR^n$.
\end{exm}
 \begin{rem}
 Our proposal for abstract {\bf interpretation scheme} of the theory looks
 as follows:\nl
\noidt The set of observables is $C_s(\overline\mSs,\mLH)$\ (cf.
Remark~\ref{rem;obs}). Their {\bf expectations} in the
states $\mu_{\{\mrh\}}$ are
\bequ
\lb\hat\ra\rb_{\{\mrh\}}\equiv\sum_j\mlam_j Tr(\mrh_j \ra(\mrh_j)).
\enqu
The higher momenta are $\lb\hat\ra^k\rb_{\{\mrh\}},\ k=2,3,\dots$, what
allows us to calculate probability distributions of (also microscopic)
observables.
The ``classical'' distribution of elementary mixtures composing a genuine
one can be given by an arbitrary (not only discrete) measure $\mu$\ on \Ss\
(endowed by a Borel structure~\cite{bona}). Then we have
\bequ\label{eq;expect}
\lb\hat\ra\rb_\mu = \int_{\mSs}Tr\left(\mrh\ra(\mrh)\right)\mu(\rd\mrh).
\enqu
The system is here described as a composition of a QM--system on \H, and a
classical system with phase space \Ss. The above considered nonlinear
dynamics of the
quantum system can be described as linear dynamics of this composed
system~\cite{bona}.
\end{rem}

\section*{Acknowledgments}
The author expresses his thanks to organizers of this Conference,
especially to Professor Doebner, for invitation and support.


\begin{thebibliography}{99}
\bibitem{peres}A. Peres: {\em Quantum Theory: Concepts and Methods}\
(Kluver Academic Publishers, Dordrecht -- Boston -- London, 1994).

\bibitem{bial&myc}\refer{J. Bialynicki--Birula and J. Mycielski}{Ann.
Phys}{100}{1976}{62}

\bibitem{weinb}\refer{S. Weinberg}{Ann. Phys.}{194}{1989}{336}

\bibitem{doeb&gold}\refer{H.--D. Doebner and G. A. Goldin}{{Phys.
Lett.} A}{162}{1992}{397}

\bibitem{gisin}\refer{N. Gisin}{Helv. Phys. Acta}{62}{1989}{363}
\refer{J. Polchinski}{Phys. Rev. Lett.}{66}{1991}{397} 
\refer{M. Czachor}{Found. Phys. Lett.}{4}{1991}{351}

\bibitem{bona} P. B\'ona: {\sl Quantum Mechanics with Mean -- Field
Back\-grounds}, Physics Preprint No. {\bf Ph10--91}, Comenius University,
Faculty of Mathematics and Physics, Bratislava, October 1991 (a revised
version: math-ph/9909022);
P. B\'ona: {\sl On Nonlinear Quantum Mechanics}, pp.
185--192 in {\em
Differential Geometry and Its Applications}, Proc. Conf. Opava
(Czechoslovakia), August 24--28, 1992\ (Silesian University, Opava, 1993).

\bibitem{bona1}\refer{P. B\'ona}{Czech. J. Phys.}{B33}{1983}{837}
\refer{A. Heslot}{Phys. Rev. D}{31}{1985}{1341}
 P. B\'ona: {\sl Classical Projections and Macroscopic Limits of
Quantum Mechanical Systems}\ (unpublished monograph, Bratislava, 1984,
revised version 1986).

\bibitem{B&L&M}P. Busch, P. Lahti and P. Mittelstaedt: {\em The Quantum
Theory of Measurement}\ (Springer, Berlin, 1991).

\bibitem{bra&rob}O. Bratteli and D.W. Robinson: {\em Operator Algebras  and
Quantum Statistical Mechanics}\ Vol.I (Springer, New York -  Heidelberg  -
Berlin, 1979).

\bibitem{czachor}
\refer{M. Czachor}{Physics Letters A}{225}{1997}{1}
\refer{M. Czachor and M. Marciniak}{Physics Letters A}{239}{1998}{353}
\refer{M. Czachor}{Phys. Rev. A}{57}{1998}{4122}
\refer{S. B. Leble and M. Czachor}{Phys. Rev. E}{58}{1998}{7091}
M. Czachor, quant-ph/9711054 (to be published in {\em Int. J. Theor.
Phys.}).
\end{thebibliography}
\end{document}